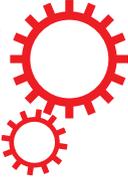

OPEN

# Cardiac kinematic parameters computed from video of *in situ* beating heart

Lorenzo Fassina[1,2,*], Giacomo Rozzi[3,7,*], Stefano Rossi[3,5], Simone Scacchi[4], Maricla Galetti[5], Francesco Paolo Lo Muzio[3], Fabrizio Del Bianco[1,2], Piero Colli Franzone[6], Giuseppe Petrilli[7], Giuseppe Faggian[7] & Michele Miragoli[3,5,8,9]

Mechanical function of the heart during open-chest cardiac surgery is exclusively monitored by echocardiographic techniques. However, little is known about local kinematics, particularly for the reperfused regions after ischemic events. We report a novel imaging modality, which extracts local and global kinematic parameters from videos of *in situ* beating hearts, displaying live video cardiograms of the contraction events. A custom algorithm tracked the movement of a video marker positioned *ad hoc* onto a selected area and analyzed, during the entire recording, the contraction trajectory, displacement, velocity, acceleration, kinetic energy and force. Moreover, global epicardial velocity and vorticity were analyzed by means of Particle Image Velocimetry tool. We validated our new technique by i) computational modeling of cardiac ischemia, ii) video recordings of ischemic/reperfused rat hearts, iii) videos of beating human hearts before and after coronary artery bypass graft, and iv) local Frank-Starling effect. In rats, we observed a decrement of kinematic parameters during acute ischemia and a significant increment in the same region after reperfusion. We detected similar behavior in operated patients. This modality adds important functional values on cardiac outcomes and supports the intervention in a contact-free and non-invasive mode. Moreover, it does not require particular operator-dependent skills.

Myocardial ischemic injury, resulting from severe impairment of the coronary blood flow, is usually related to thrombosis or other acute alterations of coronary atherosclerotic plaques[1]. Cardiovascular surgery re-establishes the blood flow in the downstream ischemic tissue by percutaneous coronary intervention (PCI)[2,3] or coronary artery bypass graft (CABG)[4]. While several imaging techniques are present for post-operative evaluation of cardiac function (e.g., CT-scan[5], MRI[6], SPECT[7]), only trans-esophageal echocardiography (TEE) can directly assist surgeons during open-chest surgery assuring, before closing the chest, an assessment of good-prognosis[8]. TEE can be combined with other imaging techniques also in the context of structural heart disease[9]. However, TEE handling requires essential skills for the operator[10] and primarily returns information only about global left ventricle function.

Almost 3% of CABG patients undergo to a second open-chest operation, mainly due to vein graft failure alone or combined with other causes such as restenosis[11], perioperative ischemia/myocardial infarction[12] and ventricular aneurisms[13]. Information about the kinematic parameters specifically from the re-vascularized cardiac regions (i.e., once the blood flow is restored after reperfusion) is missing. To date, several efforts have been devoted to measuring the kinematics of local cardiac regions following the movement of a radio-opaque marker glued/

[1]Dipartimento di Ingegneria Industriale e dell'Informazione, Università degli Studi di Pavia, Via Ferrata 1, 27100 Pavia, Italy. [2]Centre for Health Technologies (C.H.T.), Università degli Studi di Pavia, Via Ferrata 1, 27100 Pavia, Italy. [3]Dipartimento di Medicina e Chirurgia, Università degli Studi di Parma, Via Gramsci 14, 43124 Parma, Italy. [4]Dipartimento di Matematica, Università degli Studi di Milano, Via Saldini 50, 20133 Milano, Italy. [5]CERT, Centro di Eccellenza per la Ricerca Tossicologica, INAIL-exISPESL, Università degli Studi di Parma, Via Gramsci 14, 43124 Parma, Italy. [6]Dipartimento di Matematica, Università degli Studi di Pavia, Via Ferrata 1, 27100 Pavia, Italy. [7]Dipartimento di Cardiochirurgia, Università degli Studi di Verona, Ospedale Borgo Trento, P.le Stefani 1, 37126 Verona, Italy. [8]Humanitas Clinical and Research Center, Via Manzoni 56, 20090 Rozzano, Italy. [9]Institute of Genetic and Biomedical Research, National Research Council, Via Manzoni 56, 20090 Rozzano, Italy. *These authors contributed equally to this work. Correspondence and requests for materials should be addressed to M.M. (email: michele.miragoli@unipr.it)

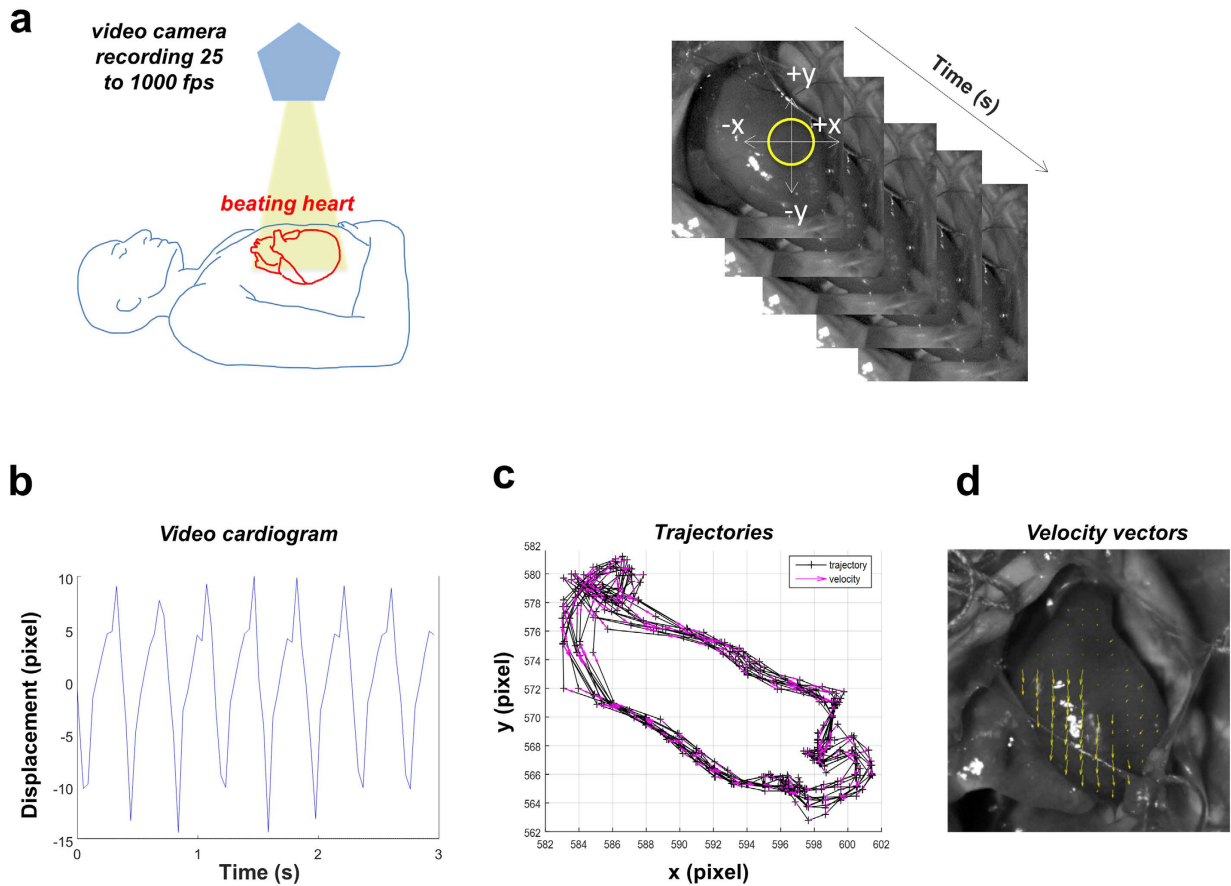

**Figure 1. Workflow with video camera positioning, video recording and evaluation of the systolic and diastolic phases.** (**a**) Left. Schematic representation of the camera positioned using an articulating arm on top of the open chest. The authors drew the scheme. Right. Sequence of video frames captured at 500 fps from a beating heart with the video marker (yellow circle) 'anchored' to the cardiac tissue while moving in x-y directions. (**b**) Video cardiogram (ViCG) showing the displacement of a selected video marker with contraction/relaxation peaks and peak-to-peak intervals. (**c**) Counterclockwise trajectories of contraction (left to right) and relaxation (right to left) for every cardiac cycle with related velocity vectors (pink arrows). The graph shows a typical overlapping in the x-y movements of the selected video marker. (**d**) Particle Image Velocimetry (PIV) showing the velocity vectors (yellow arrows) of the beating cardiac tissue.

sewed on the cardiac tissue by a biplane cineradiography[14,15]. The only reported video-tracking acquisition on the motion of a beating heart is related to mechanical compensation of a robotic arm movement during endoscopic CABG[16]. During open-chest procedures, all cardiac surgeons should confirm that a direct real-time rapid and contactless measurement of kinematic parameters, strictly related to a localized portion of the heart (i.e., ischemic or suffering regions), would be a benefit.

We show, using a commercial low-speed full HD video camera (in ~60-bpm human hearts during CABG) as well as a high-speed video camera (in ~180-bpm rat hearts that underwent ischemia/reperfusion protocol), the tangible possibility to acquire a video (from few seconds to a minute) and display a video cardiogram (ViCG) and its related kinematic parameters in a contactless modality. An algorithm runs in post-processing mode by directly displaying the trajectories of selected tissue video markers and by computing their related kinematic parameters (such as frequency, displacement, velocity, acceleration, kinetic energy and force) during cardiac cycles. From the same video the algorithm, via an adapted Particle Image Velocimetry (PIV) tool, calculates the velocity and the vorticity (i.e., the rotation frequency of the velocity vectors)[17,18] during systolic and diastolic phases. Recently, we have validated and acquired *in vitro* similar kinematic parameters in beating 2D syncytia of cultured cardiac cells[19] or in IPSCs-derived cardiomyocytes[20]. An application of *in situ* beating heart has never been proposed. Thus, the aim of this study is to introduce a contactless and user-friendly technology that can return, together with conventional gold standard techniques, important local and global functional parameters useful from experimental to clinical benches.

### Results

Both low-speed and high-speed cameras have been positioned above the heart in open-chest (Fig. 1a, left panel) Wistar rats[21] or patients, which underwent CABG[22]. The camera and the related optical macro-objective maximized the visualization of the heart, which was covered by the entire field of view.

| Parameter name | Formula | Physiological meaning |
|---|---|---|
| Displacement (pixel) | $\mathbf{U} = d\mathbf{x}$ | Marker displacement between 2 consecutive video frames: it estimates the instantaneous movement of the cardiac tissue and builds the marker trajectory. |
| Velocity (pixel/s) | $\mathbf{v} = d\mathbf{x}/dt$ | Marker velocity between 2 consecutive video frames: it estimates the instantaneous motility or contractility of the cardiac tissue and builds the marker trajectory. |
| Kinetic energy (pixel$^2$/s$^2$) | $E = \frac{1}{2}\mathbf{v}^2$ | Marker kinetic energy between 2 consecutive video frames: it estimates the consumption of ATP to generate the cardiac movement. |
| Frequency (Hz) | f = beat number/video duration | Contraction frequency calculated from the cardiac beats identified. |
| Acceleration (pixel/s$^2$) | $\mathbf{a} = d\mathbf{v}/dt$ | Marker acceleration between 2 consecutive video frames: it estimates the instantaneous variation of the motility of the cardiac tissue and builds the marker trajectory. |
| Force (N) | $\mathbf{F} = m\mathbf{a}$ | Cardiac force moving the mass m. |

**Table 1. Kinematic and dynamic parameters calculated from video marker tracking with their physiological meaning.** According to classical and Hamiltonian mechanics, **x**(t) is the position vector (pixel) in the coordinate system (x,y) of an orthonormal Euclidean space, where t is the time (s).

**Validation of the measured parameters in rat hearts.** Using Video Spot Tracker (VST, CISMM, Computer Integrated Systems for Microscopy and Manipulation, UNC Chapel Hill, NC, USA), we tracked the spatial-temporal coordinates x, y and t of the chosen video marker, in each frame (Fig. 1a, right panel)[19]. Coordinates were analyzed by a custom algorithm implemented with Matlab Programming Language (The MathWorks, Inc., Natick, MA, USA). The algorithm computed tissue displacements both in the x and y directions (Fig. 1b) and, consequently, obtained the following parameters: the contraction peaks, the relaxation peaks, the beat-to-beat intervals, the systolic and diastolic phases, and the tissue trajectory (Fig. 1c). In particular, we took in consideration the following physical quantities (Table 1):

 (i) Displacement (pixel); the instantaneous movement of the cardiac tissue in x and y directions; it builds the temporal-spatial trajectory.
 (ii) Velocity (pixel/s); the instantaneous motility or contractility of the cardiac tissue; it builds the temporal-spatial trajectory.
(iii) Kinetic energy (pixel$^2$/s$^2$); an estimation of the consumption of ATP for the generation of cardiac movement.
(iv) Frequency (Hz); contraction frequency calculated from the identified cardiac beats.
(v) Acceleration (pixel/s$^2$); the instantaneous variation of the cardiac tissue motility; it builds the temporal-spatial trajectory.
(vi) Force (N); cardiac force moving a mass.

The calculus of the preceding quantities was subjected to a de-noising procedure via wavelet compression (near symmetric wavelet: Symlets 4; decomposition level: 3; compression method: global threshold leading to recover 99% of the signal energy).

To study the cardiac cycle, we obtained fields of velocity vectors (Fig. 1d) in diastole and in systole by means of PIV. Consequently, it was possible to evaluate the mean rotation frequency of the velocity vectors (i.e., the mean vorticity in Hz) and the mean velocity module (calculated with respect to space and time).

*Temporal and spatial resolution.* During the initial setting, we studied the effect of both video marker radius and rate effect ('frame number per second' or fps) of the video acquisition on the returned data quality. In Supplementary Fig. 1, we reported the temporal graphs of the velocity marker in the x direction (the y direction produced similar results and therefore it is not shown) in normal (HEALTH) heart. The radius equal to 15 pixel was the minimum required to correctly follow ventricular epicardial surface movements and, then, to correctly identify cardiac beats.

In Supplementary Fig. 2a–f, we displayed trajectory and velocity vectors, which were obtained at different acquisition rates (500, 250, 125, 100, 50, and 25 fps). As expected, the rate of 500 fps gave the best trajectory and velocity results (Supplementary Fig. 2a). The rates between 250 and 30 fps could be considered acceptable (Supplementary Fig. 2b–e). However lower rates (≤25 fps, Supplementary Fig. 2f) were significantly affected by the aliasing phenomenon with loss of trajectory and velocity details and thereby not used in the present study.

*Reproducibility of the acquired data.* We evaluated the aforementioned parameters during time by acquiring 1-s video recording of sinus rhythm at the highest possible temporal resolution (1 kHz) onto three healthy rat hearts, every 10 min for 1 h (Supplementary Fig. 3, Supplementary Video 1). We anchored the video marker in the same position and monitored changes in displacement, velocity, acceleration and energy over time. We observed minimal percentage variations (likewise physiological) for the four parameters, suggesting a good reproducibility of our data.

*Kinetic energy acquisition in a controlled and simple system.* We then evaluated the kinetic energy parameter by acquiring videos of the periodic motion of a pendulum (mass of 200 g) oscillating for small displacements (Supplementary Fig. 4a). The pendulum was connected to an arm by a 21.5 cm not-stretchable string and the motion was acquired at 200 fps for five seconds. We selected the video marker onto the white area of the pendulum and displayed the trajectories (Supplementary Fig. 4b), the coordinate x vs. time (Supplementary Fig. 4c) and the kinetic energy at the highest, intermediate and lowest positions (Supplementary Fig. 4d). As expected, marker movement is an arc. When the coordinate x was displayed vs. time, it drew a periodic wave. The kinetic energy we measured for five consecutive frames in the three aforementioned positions was ca. zero at the highest position (all the energy is potential; h = 17 cm) and maximal at the lowest position (all the energy is kinetic; $6.49 \times 10^6 \pm 6.01 \times 10^5$ pixel$^2$/s$^2$, h = 11.5 cm). In the intermediate position, the kinetic energy was $4.26 \times 10^6 \pm 5.40 \times 10^5$ pixel$^2$/s$^2$.

*Measurement of the contraction force.* We then measured the ventricular force by anchoring a known spherical mass (0.035 g, diameter = 3 mm) onto epicardial surface in three different rat hearts (Fig. 2a, Supplementary Video 2). The use of a mass connected to the cardiac tissue allowed a switch from a description in terms of acceleration (pixel/s$^2$) to the International System of Units and, consequently, we calculated the contraction force (N) accelerating the mass (Fig. 2b) and the related ViCG in terms of displacement and velocity (Fig. 2c). The ViCG was also useful to recognize the diastolic and systolic phases.

We found the following mean contraction forces (Fig. 2b): i) $3.29 \times 10^{-5} \pm 5.62 \times 10^{-6}$ N (n of beats = 72); ii) $3.30 \times 10^{-5} \pm 9.09 \times 10^{-6}$ N (n of beats = 31); and iii) $3.86 \times 10^{-5} \pm 6.72 \times 10^{-6}$ N (n of beats = 40), suggesting that every cardiac cycle exerts a well-conserved force of contraction.

We also observed, for every beat, a well-conserved contraction-relaxation trajectory, which was a kind of hysteresis loop due to the intrinsic viscoelastic properties of the myocardium[23].

### Simulation of kinematic parameter measurement in ischemic contractile hearts.

We evaluated the effectiveness of our algorithm on data obtained from numerical simulations based on the cardiac electro-mechanical coupling model. We incorporated a transmural ischemic region of 1 cm × 1 cm. The excitation process was initiated by stimulating three endocardial anterior apical sites and one endocardial posterior apical site, mimicking an idealized Purkinje network. Two simulations were performed, one for a healthy tissue (HEALTH) and one for a tissue with a transmural ischemic region (ISCH). In both cases three beats were simulated, at a basic cycle length of 500 ms. Supplementary Fig. 5 displays the epicardial distributions of transmembrane and extracellular potentials computed in the HEALTH and ISCH simulations, during the plateau phase of the heartbeat (ST segment, t = 120 ms). The ISCH transmembrane potential distribution showed a minimum in the ischemic region, whereas the ISCH extracellular potential distribution showed a maximum (ST elevation), in correspondence of the ischemic region. The kinematic evaluation applied to the experimental data was validated on the simulated data related to the healthy and ischemic tissue. The results reported in Fig. 3 show that ischemia reduced of about 20%: i) the maximum displacement module (Fig. 3a), ii) the mechanical behavior of the cardiac tissue, with a reduction of the maximum velocity module (Fig. 3b) and iii) the mean kinetic energy (Fig. 3c). These theoretical findings validated the use of kinematic parameters as indices of cardiac mechanical performance.

### Video kinematic parameters in ischemic and reperfused rat hearts.

We then acquired videos of six *in situ* rat hearts before an acute cardiac ischemia/reperfusion injury model[24] with a high-speed camera (acquisition rate = 500 fps) for 1 s in control (HEALTH), ischemic (ISCH, 6 min) and reperfused (REP, 6 min) conditions.

Figure 4 showed the data only at 6 min following reperfusion as the data at 12 min were comparable. We selected always the same video marker on the well-defined region beneath the ligation (Fig. 4b, left panel). We only extracted data where the ischemic tissue turned from deep red to trans-lucid white, as a sign of a nearly complete coronary occlusion. Alike the simulation results, the system captured that the brief ischemia patently corrupted the mechanical performance of the ischemic region in terms of trajectories, ViCGs (Fig. 4a) and kinematic parameters (Fig. 4b, right panel), while the reperfusion caused significant improvements of those parameters.

We found that the maximum velocity module was significantly reduced during ischemia from $9.71 \pm 3.09$ pixel/ms to $3.32 \pm 1.05$ pixel/ms ($p < 0.05$ vs. healthy tissue), while, after 6 min of reperfusion, this parameter showed significantly higher and more physiological values ($5.88 \pm 1.95$ pixel/ms, $p < 0.05$ vs. ischemic tissue, $p < 0.05$ vs. healthy tissue).

We observed an analogous behavior for the mean kinetic energy. Specifically, it significantly decreased during ischemia from $1.60 \pm 0.68$ pixel$^2$/ms$^2$ to $0.27 \pm 0.11$ pixel$^2$/ms$^2$ ($p < 0.05$ vs. healthy tissue), whereas, after 6 min of reperfusion, this parameter showed significantly higher and more physiological values ($0.68 \pm 0.16$ pixel$^2$/ms$^2$, $p < 0.05$ vs. ischemic tissue, $p < 0.05$ vs. healthy tissue).

The mean acceleration module significantly decreased during ischemia from $1.51 \pm 0.84$ pixel/ms$^2$ to $0.63 \pm 0.16$ pixel/ms$^2$ ($p < 0.05$ vs. healthy tissue), whereas, after 6 min of reperfusion, this parameter showed physiological and significantly higher values ($1.14 \pm 0.22$ pixel/ms$^2$, $p < 0.05$ vs. ischemic tissue, not significant vs. healthy tissue). In addition, the mean frequency of contraction was also decreased in the ischemic heart in comparison with the healthy one (from $4.08 \pm 1.81$ Hz to $3.64 \pm 0.45$ Hz, $p > 0.05$), whereas the reperfusion did not change the contraction frequency in comparison with ischemia ($p = 0.21$). The same experiment was performed for the non-ischemic regions (right ventricles) in the HEALTH, ISCH and REP conditions and we did not find significant differences for the aforementioned parameters (data not shown).

During the same protocol, PIV was used to study some physical parameters of the cardiac cycle. In particular, as example of PIV imaging, we have reported the velocity vector field (pixel/s) in diastole (Fig. 5a, left panels) and in systole (Fig. 5a, right panels).

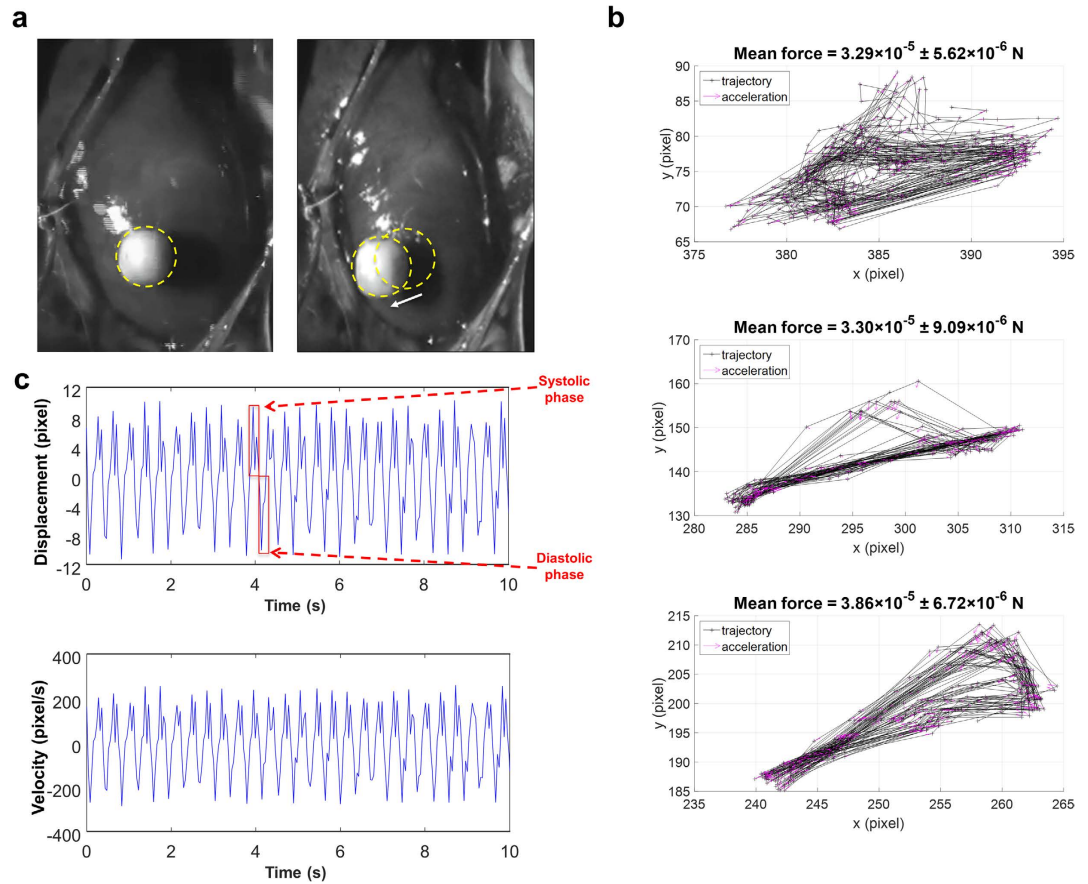

**Figure 2. Validation of contraction force measured with a known spherical mass anchored onto the epicardial surface. (a)** Position of the spherical mass at the start of systole (left) and at the start of diastole (right) with the related displacement vector (white arrow) between the two preceding moments. **(b)** Trajectory and acceleration of the mass in three different rat hearts with the mean contraction force (N) during the observation period (10 s). Data are expressed as mean ± SEM. **(c)** Video cardiograms of displacement vs. time (top) and of velocity vs. time (bottom). Systolic and diastolic phases are underlined in red.

The mean velocity module was decreased in ischemia in comparison with the healthy heart (from $43.10 \pm 6.56$ pixel/s to $36.24 \pm 4.18$ pixel/s, $p > 0.05$, Fig. 5b); the reperfusion ameliorated, but not significantly, the mean velocity module (to $41.19 \pm 5.23$ pixel/s) in comparison to ischemia. Compared with the initial value measured for the healthy heart, mean velocity module remained not significant ($p > 0.05$).

The mean vorticity was significantly decreased in the ischemic heart in comparison with the healthy one (from $0.52 \pm 0.18$ Hz to $0.31 \pm 0.05$ Hz, $p < 0.05$, Fig. 5c); the reperfusion ameliorated the mean vorticity (to $0.33 \pm 0.06$ Hz) in comparison to ischemia ($p > 0.05$) but we did not obtain the initial value of the healthy heart ($p < 0.05$).

**Video kinematic parameters in patients underwent CABG.** We then moved to experimental medicine by adapting our system to an operating room with the purpose of recording the beating hearts of ten patients that underwent CABG (Fig. 6).

In average, the coronary occlusion for ten patients was $86.1 \pm 2.5\%$. For all hearts, we selected a video marker (radius = 20 pixel) onto the epicardial surface beneath the region where CABG was performed. We performed three video recordings for each patient: i) at chest opening (at pericardium opening, time 0 min); ii) after CABG (but before protamine sulfate infusion with cannulated aorta, time 120–180 min); iii) before chest closing (time 140–200 min). The TEE showed a significant improvement for the mean ejection fraction (EF%, from $31.75 \pm 4.80$ to $46.70 \pm 2.60$, $p = 0.004$) and a not significant increment for the mean fractional shortening (FS%, from $38.40 \pm 6.10$ to $50.27 \pm 2.50$, $p = 0.08$) (Supplementary Fig. 6a). With our method, we did observe significant changes for the acceleration module ($p = 0.02$, Fig. 6a, Supplementary Fig. 6b) and a substantial improvement for the other three parameters: maximum velocity module ($p = 0.28$), kinetic energy ($p = 0.58$), and maximum displacement module ($p = 0.37$) (Fig. 6a, Supplementary Fig. 6b). However, when we looked at the individual patient's outcome, patient #3, despite increased acceleration module, denoted a substantial reduction in the other three parameters.

The corresponding ViCGs showed that cardiac cycles differed from each other for all patients before CABG. For the sake of comparison, we displayed the ViCG from patient #1 (which obtained an increment in all preceding parameters after surgery) (Fig. 6b, upper panel) and patient #3 (Fig. 6c, upper panel) before CABG. In both patients, we observed disturbed cardiac cycles and dissimilar contraction/relaxation trajectories (data not shown). After CABG, the cardiac cycles became more physiological for patient #1 (Fig. 6b, lower panel) but not for patient #3 (Fig. 6c, lower panel).

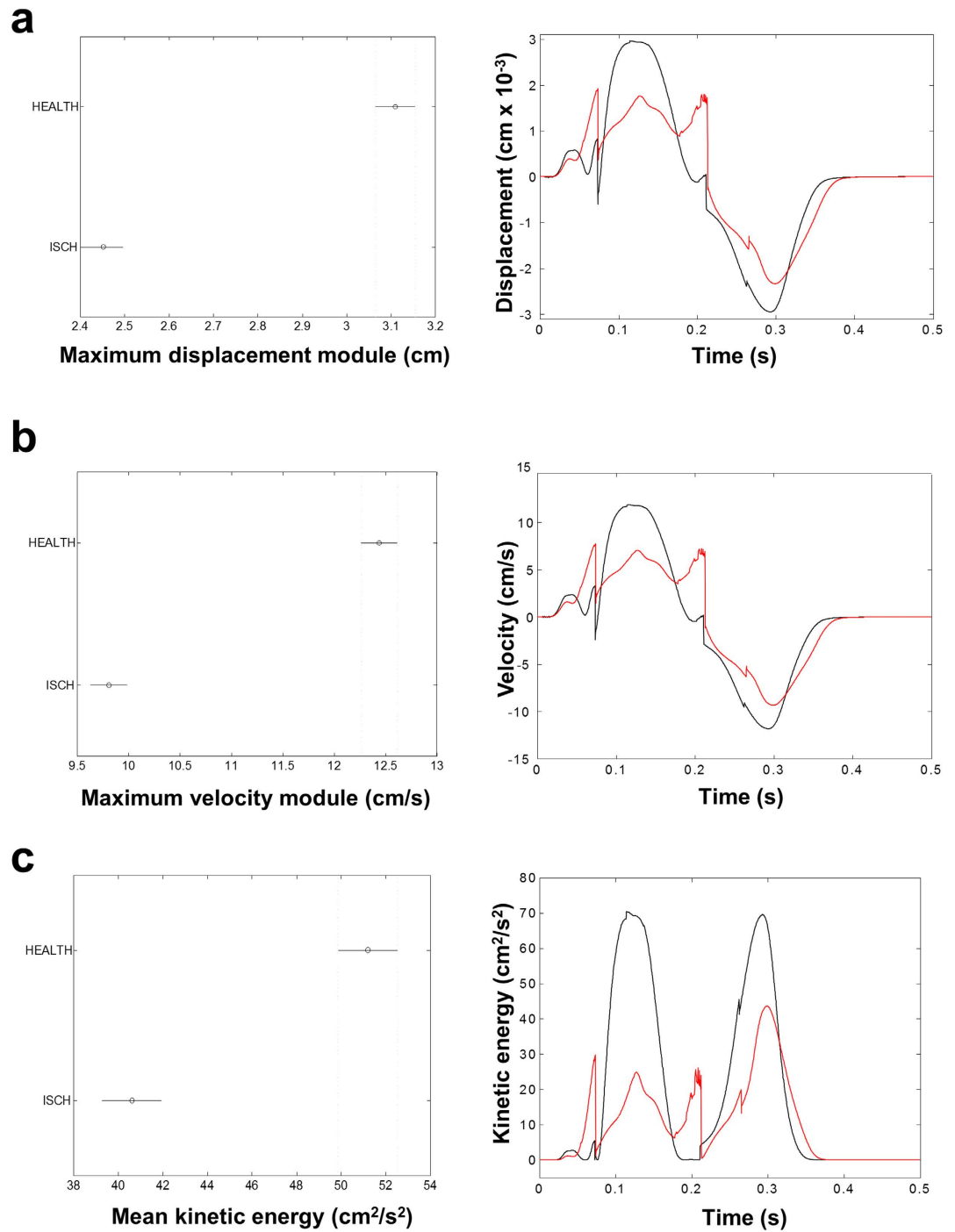

**Figure 3. Simulation of kinematics in ischemic heart. (a)** Left. Maximum displacement module, as defined in Table 1, in marker trajectories (frame distance = 0.25 ms), which was averaged across 48 healthy (HEALTH) and 48 ischemic (ISCH) sites. The horizontal bars are the 95% confidence intervals for the differences between means according to one-way ANOVA and LSD (Least Significant Difference) *post hoc* test. There is a statistically significant difference between the means with non-overlapping bars ($p < 0.05$). Right. Time evolution of the displacement of a single sample site; comparison between healthy (black) and ischemic (red) site. **(b)** Same as (**a**) for the maximum velocity module. **(c)** Same as (**a**) for the mean kinetic energy.

Similarly, PIV analysis, before and after CABG in diastolic and systolic phases, for aforementioned patients, showed that the mean velocity module significantly increased by 32% ($p < 0.05$) in patient #1 (Fig. 7c) and only by 9% in patient #3 (not significant, Fig. 7d). Accordingly, patient #3 received an inotropic agent and extra monitoring in the critical-care postoperative unit.

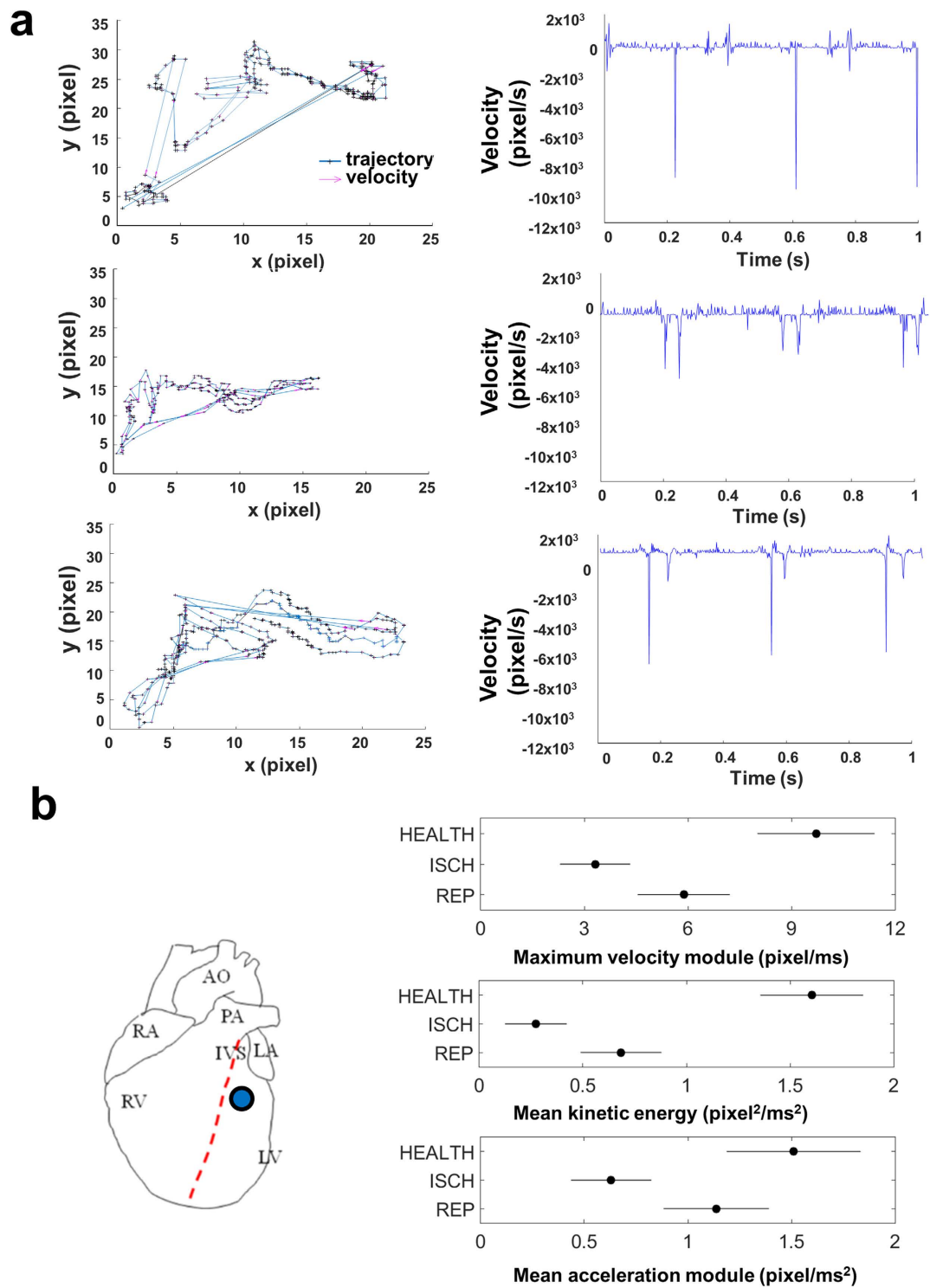

**Figure 4. Kinematic parameters and ViCGs obtained by using a high-speed video camera during ischemia/reperfusion protocol in rats.** (**a**) Left column. Trajectories and velocities measured from a 1-s recording at 500 fps during ischemia/reperfusion protocol. Top: healthy tissue. Middle: ischemic tissue, 6 min after coronary ligation. Bottom: reperfused tissue, 6 min after reperfusion. Right column. Video cardiogram of velocity vs. time in the three conditions: healthy (top), ischemic (middle) and reperfused (bottom) tissue. (**b**) Left. Representative video marker position (blue circle) on a schematic rat heart. A video marker with a radius of 20 pixel was selected onto the tissue underneath the coronary ligation. Right. Top. Maximum velocity module measured in healthy (HEALTH), ischemic (ISCH) and reperfused (REP) tissue. Middle. Same for mean kinetic energy. Bottom. Same for mean acceleration module. See the text for statistical significances. The horizontal bars are the 95% confidence intervals for the differences between means according to one-way ANOVA and LSD (Least Significant Difference) *post hoc* test. There is a statistically significant difference between the means with non-overlapping bars ($p < 0.05$). n = 6.

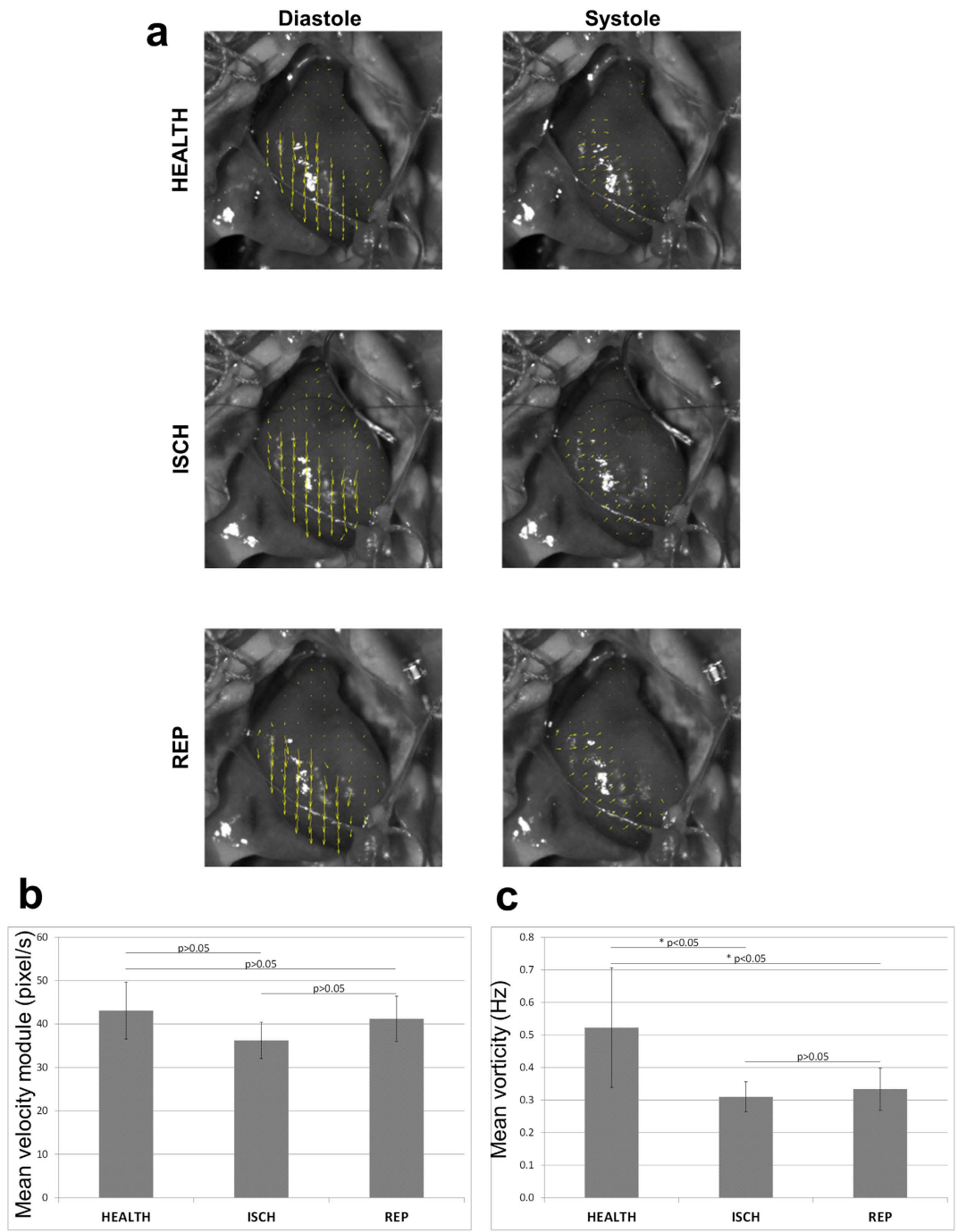

**Figure 5. Particle Image Velocimetry (PIV) analysis in ischemic/reperfused rat hearts.** (**a**) PIV showing the velocity vectors for a rat heart as example in diastole (left column) and in systole (right column) for the HEALTH, ISCH and REP conditions. (**b**) Velocity module calculated by PIV analysis in the three aforementioned conditions. (**c**) Same as (**b**) for vorticity. The p-value was *$p < 0.05$, calculated from Student's t test (paired). n = 6.

**Improved kinetic energy following atrioventricular block in rat and human hearts: the Frank-Starling effect.** We ran a protocol whereby a gentle air-pressure mechanical stimulation (20 kPa for 1 s), delivered onto the rat atrioventricular (AV) regions, exerted a limited stretch-induced depolarization and caused AV blocks, resembling the condition of the *commotio cordis*[16,25,26]. This intervention delayed the preload and resulted in an increment of the end-diastolic volume. Following the Frank-Starling mechanisms, the stroke volume increased and therefore the required kinetic energy as well (Supplementary Video 3, Supplementary Fig. 7a–b)[27]. Our methodology detected, following the AV block, an increased kinetic energy from 0.003 to 0.025 pixel$^2$/ms$^2$.

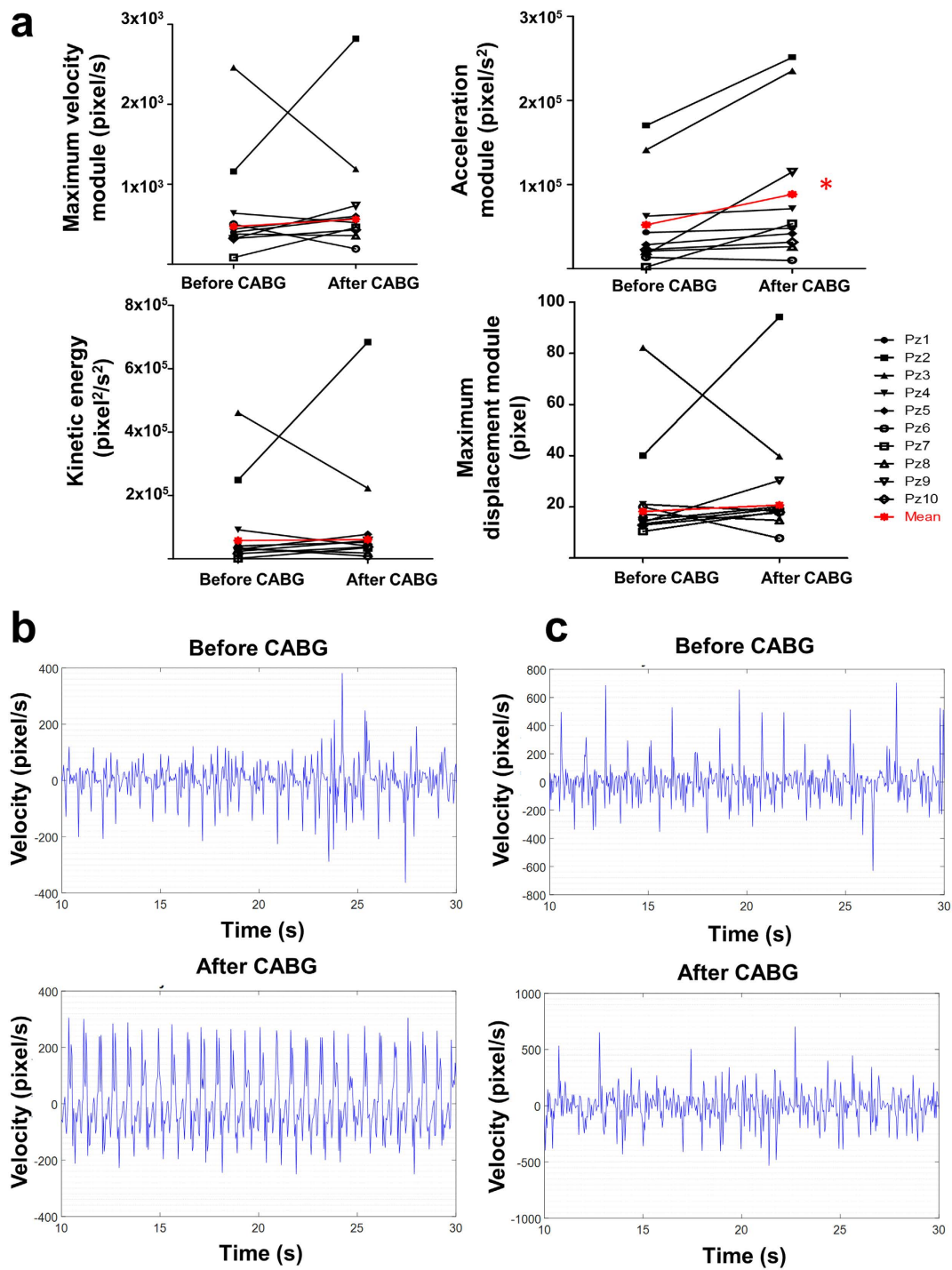

**Figure 6. Kinematic parameters and video cardiograms in patients underwent coronary artery bypass graft (CABG). (a)** Top. Maximum velocity module and acceleration module measured by using a low-speed video camera. Bottom. Same for kinetic energy and maximum displacement module. (**b**) Video cardiogram (velocity vs. time) for the patient #1 before (top) and after (bottom) CABG. (**c**) Same as (**b**) for patient #3.

Similarly, we found a spontaneous increment of the end-diastolic volume in human heart (patient #2, Supplementary Fig. 7c–d) after CABG intervention but before protamine sulfate infusion. We observed an increment in kinetic energy after the pause from 0.084 to 0.149 pixel$^2$/ms$^2$ before returning to 0.093 pixel$^2$/ms$^2$.

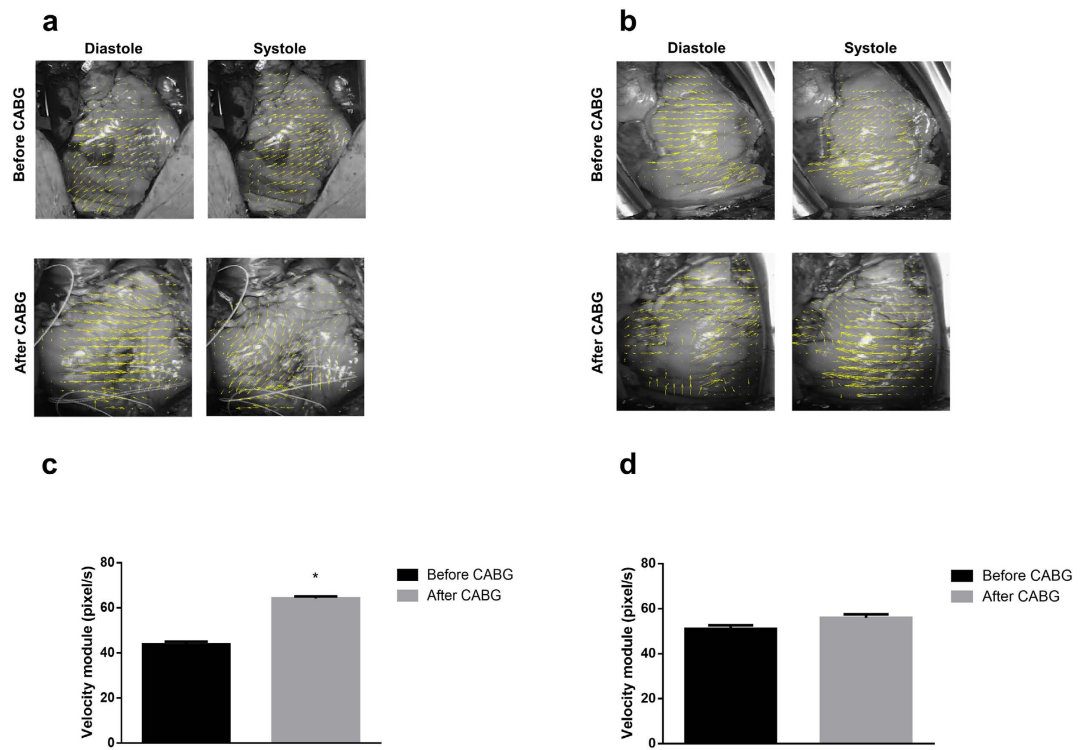

**Figure 7. Particle Image Velocimetry (PIV) analysis in CABG patients.** (**a**) PIV analysis for patient #1 before (top) and after (bottom) CABG for the diastolic (left) and systolic (right) phases. (**b**) Same as (**a**) for patient #3. (**c**) Velocity module calculated by PIV analysis for patient #1 before and after CABG (*p < 0.05). (**d**) Same as (**c**) for patient #3. Calculated from one-way ANOVA, LSD *post hoc* test. n = 1,732 video frames for patient #1 and n = 992 video frames for patient #3.

## Discussion

We report a video-based and contactless technique that can add, for the first time to our knowledge, precise cardiac kinematic information during open-chest procedures. Our technology can be adapted for both basic and clinical research and can work in conjunction with standard monitor procedures, without disturbing the intervention during open-chest cardiac surgery. We validated our kinematic parameters in the 'physical' case, i.e., tracking the movement of a pendulum during its periodic motion and in rats by connecting a known mass onto the epicardial surface and following its trajectories. The minimal differences we encountered in the force or acceleration measurements were only related to physiological changes in the kinematics of the cardiac tissue and were not an artifact of the measure (cf. Fig. 2). We also demonstrated the reproducibility of the data by monitoring over time our kinematic parameters from the same hearts (Supplementary Fig. 3). We were able to capture the contraction movement of the ischemic/reperfused tissue area and we computed the corresponding kinematic and dynamic parameters (such as displacement, velocity, kinetic energy, frequency, acceleration, force) which are undetectable via conventional TEE. We precisely interrogated the same epicardial areas before and after intervention i) *in silico* by modeling ischemic contractile hearts, ii) in an experimental rat model during ischemia/reperfusion protocol, and iii) in patients underwent CABG.

Our image processing technology allowed monitoring the progression of the intervention by analyzing the cardiac kinematic parameters (by video markers with a radius of at least 15–20 pixel and a temporal resolution of ≥ 30 fps, Supplementary Figs 1–2). These data can be extracted and presented in the operating room before closing the patient's chest, permitting supplementary considerations about cardiac performance and prognosis. While the human heart beats at an intrinsic frequency of ca. 1.5 Hz, rat heart beats at ca. 4.5 Hz. Thus, we decided to use either a low-cost camera for human studies (full HD portable camcorder, built-in memory) with auto-zoom and low temporal resolution or a high-speed camera for animal studies (1000 fps with manual zoom, video acquisition connected to a fast PC via frame grabber). In all experiments, we never lost video marker tracking. We obtained data on patients before and after CABG (at the final check and after protamine sulfate infusion) showing that the acceleration of the reperfused tissue significantly increased (Fig. 6a). To note, also the other three parameters showed an increment in average.

The differences observed in rats vs. humans (i.e., statistical significance was achieved in all ischemic rat parameters) were possibly related to the artificial almost-complete coronary occlusion in healthy rat hearts vs. pathophysiological coronary occlusion of 86.1% occurring in the failing human hearts (inclusion selection for CABG vs. stent placement).

For human hearts, the local cardiac kinematics before and after CABG was extremely variable (cf. Fig. 6), possibly depending on the volume of ischemic region, age, gender, weight, lifestyle and medical records (i.e., when coronary occlusions have been detected). Moreover, we have to take into consideration that the field of view of open-chest human heart covers mainly the right ventricle compared to the rat heart where the left ventricular (LV) area, within the field of view, is larger.

There is clinical evidence supporting the concept that all patients undergoing CABG have different degrees of myocardial stunning[28], which occasionally requires inotropic support after surgery (i.e., patient #3), but not signs of myocardial infarction.

In animals, stunned myocardium has been detected after 15–30 min of coronary occlusion[29], which was not our case. Certainly, prolonged ischemia would be important for understanding whether the method can detect abrupt evidence of myocardial pathophysiological alterations[30]. Indeed pioneering studies indicate that, following brief episodes of ischemia, creatine phosphate[31] increases rapidly after reperfusion, suggesting that local kinetic energy rapidly increases as well[32].

The power of our technique is that the ViCG, which was obtained during ischemia/reperfusion protocol or before and after CABG, may return an overall view of the quality of the intervention from a mechanical perspective, especially in the reperfused regions, where the video markers were positioned. In addition, the ViCG can be combined with the conventional ECG by returning information on *in vivo* data of excitation-contraction coupling machinery.

The benefit of using high-fps acquisition in basic research (up to 1000 fps) is such that we can distinguish local contraction trajectories in a millimeter range during ventricular fibrillation (data not shown) in rat hearts or we can appreciate an increment of the kinetic energy after a mechanical stimulus inducing AV blocks (cf. Supplementary Fig. 7). The technique can be easily adapted to study other inherited or acquired cardiac complications at both laboratory and clinical levels by returning important physiological data on tissue contraction at open-chest stage.

In summary, our technology can monitor and examine several cardiac regions in open-chest beating hearts. This not only provides mechanical insights, but it also offers new prognostic values. Moreover, kinetic energy, as indirect measurement of ATP consumption[19], can be embraced as a clinical marker.

**Limitations.** Video kinematic evaluation performed from a single camera is not able to study the movement in the z direction thus resulting in bi-dimensional analysis of three-dimensional (3D) phenomena. Nevertheless, the designed and implemented algorithm is already capable to acquire 3D motion and it was employed to study the 3D data of mathematical modeling. A further development for a stereoscopic/double camera acquisition as well as concurrent monitoring of electrical activity could solve the limitations. Acceleration and kinetic energy are obtained from the linear parameters (displacement and velocity), so they are *per se* prone to the measure noise. We compared our data from the same sampling rate, i.e., HEALTH vs. ISCH in animal studies or before and after CABG in human studies. Doing so, we avoided the aliasing phenomena by maintaining the same signal-to-noise ratio.

While we could measure kinematics on the left ventricle in animal studies, only the right ventricle is exposed in humans, thereby limiting our measurement at the closest regions near the CABG intervention.

## Methods

**Experimental animals.** The study population consisted of nine male Wistar rats bred in the animal facility of the Department of Life Science, University of Parma. The rats were 12–14 weeks of age and weighed 300–350 g. The investigation was approved by the Veterinary Animal Care and Use Committee of the University of Parma (Italy) (Protocol # 41/2009 and 59/2012) and it conformed to the National Ethical Guidelines of the Italian Ministry of Health and to the Guide for the Care and Use of Laboratory Animals (National Institute of Health, Bethesda, MD, USA, revised 1996). More information is detailed in the Supplementary methods section.

**Human patients.** We performed video recording in patients during CABG surgery in the Azienda Ospedaliera Universitaria Integrata (Verona, Italy). Ten patients (9 males, 1 female) were enrolled from March to December 2016. Patients were selected for CABG as they presented coronary occlusion of 86.1% in average. No human tissue samples have been used. The video recording never interfered with the standard CABG intervention, in accord with the guidelines for coronary artery bypass surgery[26]. All methods and experiments were carried out in accordance with relevant guidelines and regulations set by the Institutional Review Board (IRB) of the Azienda Ospedaliera Universitaria Integrata of Verona, where all experiments were done (approved protocol from ethics committee on 16th March 2016, # 847CESC Protocol # 13371). All patients signed an informed consent.

**Functional imaging.** *Low-speed video camera.* A full HD (1920 × 1080 total pixel area) camera Samsung S10 (with an internal SD memory of 256 MB and internal rechargeable battery) has been used for the cardiac force measurement in rats and for the CABG recording in the operating room. The on-site monitor of the camera allowed the recording of the heart beating. Videos have been recorded for 60 s in full HD gray-scale mode at 30 fps. The hearts have been constantly illuminated with scialytic light by keeping it in the same position during video recording for both animals and patients.

*High-speed video camera.* A Baumer HXC13 (Baumer Italia, S.r.l., Milano, Italy) camera with full CameraLink® interface (1280 × 1024 total pixel area for 500 fps or 1020 × 600 for 1000 fps) was used for ischemia/reperfusion experiments in rats. The camera was equipped with a macro-objective Kowa Industrial Lenses LM35XC, F = 1:2.0, f = 35 mm, picture size 13.8–18.4 mm (RMA Electronics, Hingham, MA, USA). The acquisition software was custom made in LabVIEW Visual Programming Language (National Instruments, Assago, Milano,

Italy). We ran 1-s acquisitions at either 1000 fps (reproducibility data) or 500 fps (ischemia/reperfusion data) allowing the recording of spontaneous beats in anesthetized rats. No pixel binning was used. The camera was connected with two CameraLink® cables to a frame grabber acquisition board PCIe 1433 (National Instruments, Italy) adapted into a Workstation HP Z220 (Crisel Instruments, Italy) with 24 GB RAM. The video file consisted of 1000 or 500 TIFF images (1.25 MB each) of the beating heart. For the pendulum experiment, we ran an acquisition at 200 fps (1280 × 1024 total pixel area) for five seconds of recording.

**Video acquisition.** *Video kinematic evaluation in rats.* In the anesthetized animals subjected to artificial ventilation (Rodent Ventilator, Ugo Basile, Italy), the heart was exposed through a longitudinal sternotomy and suspended in a pericardial cradle. Body temperature was maintained with infrared lamp radiation. In the present study, the camera was levelled at 17 cm above the spontaneously beating heart. A constant illumination was achieved using a scialytic lamp for laboratory (ElettroMedica, Italy). During recording, the ventilator was switched off to avoid motion artifacts and immediately re-switched on at the end of the recording. The distance between the camera and the heart, the focus and the heart's orientation were never changed during the experiments.

For the force measurement experiments, we employed known masses consisting of a pin-head with a 1 mm-long needle inserted on the epicardium surface. For the ischemia/reperfusion experiments, we recorded the beating heart in normal conditions followed by anterior descending coronary ligation, as recently described[33]. We left the surgical needle within the performed knot as this allowed the removal of the ligation without damaging the tissue. Then, we waited for 6 min and 12 min before recording another video of the beating heart (now ischemic). The same procedure was followed after the removal of the ligation (i.e., two video files were recorded after 6 min and 12 min).

For the mechanically-induced AV blocks, we adopted a flexible cannula (0.8 mm diameter) firmly fixed to a three-axes manipulator and positioned ca. 5 mm above the pulmonary cone. A gentle air-pressure flow (20 kPa) was delivered for 1 s using an electro-valve (Asco Numatics, Bussero, Milano, Italy) remotely controlled by a stimulator (Crescent Electronics, Sandy, UT, USA), similar to what we had previously implemented for single cell mechanical stimulation[33].

*Video kinematic evaluation during CABG in human patients.* In the operating room, the camera was positioned with an articulating arm at 40 cm above the open chests. We performed video recordings of beating hearts at chest opening (after pericardium opening, time 0 min), after CABG but before protamine sulfate infusion with cannulated aorta (data not shown, time 120–180 min), and before chest closing (time 140–200 min). All patients were in normal sinus rhythm during the intervention. Because, during CABG, the surgeons frequently move the operating table, we recorded the original position of the table and recalled it during video acquisitions. For the analysis (before and after CABG), we took into consideration additional anatomical markers, such as coronary vessels or skin edges for properly positioning the video marker.

**Quantitative analysis.** *Tracking of the cardiac contraction.* Using VST program (http://cismm.cs.unc.edu/downloads), which tracks the movement of biological objects starting from the video recording of their displacements, we anchored a single video marker to the cardiac tissue or to a mass (0.035 g) firmly connected to that tissue. The x and y coordinates are expressed in pixel, whereas the t coordinate in second (s).

*Kinematics.* Using a custom algorithm based on the Matlab Programming Language (The MathWorks, Inc., Natick, MA) and the tracked spatial-temporal coordinates of the cardiac contraction, we identified the heartbeats and then calculated the corresponding kinematic parameters (see Table 1 for the mathematical definitions and their physiological meaning).

*Particle Image Velocimetry.* We employed 'Particle Image Velocimetry' (PIV) tool, which is a fluid flow visualization and quantification technique, in particular its Matlab Central implementation in PIVlab (http://it.mathworks.com/matlabcentral/fileexchange/27659-pivlab-time-resolved-particle-image-velocimetry-piv-tool?s_tid=srch-title). We applied the PIV to study the cardiac cycle and we reported the images of the velocity vectors (pixel/s) in diastole and in systole. During the whole observation period, it was possible to evaluate the mean rotation frequency of the velocity vectors, that is, the mean vorticity (Hz) and the mean velocity module (pixel/s).

**Mathematical models.** *Numerical simulations.* The numerical simulations were based on the cardiac electro-mechanical coupling (EMC) model. This model described the interplay between the spread of electrical excitation in the cardiac tissue and the consequent contraction/relaxation process. The EMC model consisted of four sub-components, two for the bioelectrical activity and two for the mechanical response of the cardiac tissue. The bioelectrical components were the Bidomain model for the electric current flow[34] and the ten Tusscher human ventricular model[35] for the cellular membrane dynamics. We disregarded the presence of an ischemic damage in the surviving layers at the border of the ischemic region. From the bioelectrical point of view, ischemic conditions were modeled modifying the following parameters of the ten Tusscher model[36]: the extracellular potassium concentration was increased from 5.4 mM to 8 mM; the maximal conductances of the $I_{Na}$ and $I_{CaL}$ currents were decreased by 25%; the parameters of the $I_{KATP}$ current were modified as in[35] for ischemia stage 2. In the normal ventricular tissue, the same set of Bidomain anisotropic conductivities was assigned, yielding the physiological conduction velocities of about 0.065 and 0.03 cm/ms, when measured for wave front propagation along and across fiber, respectively, matching reported conduction velocities. Since we considered the early stage of myocardial ischemia, the conductivity coefficients inside the ischemic region were not modified, reflecting an unchanged distribution of connexin 43. From the mechanical point of view, ischemic conditions were modeled by reducing the active tension and passive tissue stiffness of 10% and 25% with respect to their normal values. Stretch-activated channels were incorporated in the membrane model according to[37]. The macroscopic description of the mechanical activity was based on the equations of finite elasticity, where the passive mechanical properties of the myocardium were assumed to be transversely isotropic, hyper-elastic, and nearly incompressible, defined by an exponential strain energy function derived from[38]. Realistic fiber orientations were incorporated into both the conductivity tensors of the Bidomain model and the strain energy function. The EMC model was finally closed by the active tension generation model developed describing the process of calcium binding to troponin C and cross-bridge cycling triggered by calcium release from the intracellular calcium stores during the electrical activation of a myocyte. The active tension generated by this model entered as input in the non-linear elasticity equations yielding the active component of the stress tensor. The intracavitary blood pressure p in the left ventricular (LV) cavity was described by a pressure-volume loop model[39], based on the following four phases:

1. Isovolumetric LV contraction phase, where p increased from the end-diastolic pressure (EDP) value of about 2 kPa to 10 kPa;
2. Ejection phase, where the pressure-volume relationship was described by a two-element Windkessel model, until the volume reduction stopped;
3. Isovolumetric LV relaxation phase, where p decreased to 1 kPa;
4. Filling phase, where p increased linearly to EDP.

*Numerical methods.* The space discretization was based on Q1 finite elements. The computational domain was a truncated ellipsoid representing the left ventricle geometry. The size of the major axis of the ellipsoid was 5 cm, while that of the minor axis was 2.7 cm. Two different meshes were used, a fine one to solve the Bidomain equations and a coarser one for the finite elasticity equations. The nodes of the electrical mesh were 3,631,488, while those of the mechanical mesh were 8,400. The time discretization was performed by a semi-implicit finite difference scheme. The time step size was 0.05 ms for the electrical components of EMC model and 0.25 ms for the mechanical components. Our parallel code was based on the PETSc library (https://www.mcs.anl.gov/petsc/). The simulations were run on 256 cores of the Bluegene cluster Fermi of the CINECA laboratory (http://www.cineca.it). For further details on the numerical methods, we remand to[34,40,41]. Parameters calibrations are detailed in the Supplementary methods section.

**General statistics.** Normal distribution of variables was checked by the Kolmogorov-Smirnov test. Statistics of normally distributed variables included mean ± standard error of the mean (SEM), paired and unpaired Student's t test. Statistical significance was set at $p < 0.05$. For the ischemia/reperfusion experiments, one-way Analysis of Variance (ANOVA) was applied with *post hoc* Least Significant Difference (LSD) test, electing a significance level of 5% (the results are expressed as mean ± 95% confidence intervals for the differences between means). Statistics was performed using GraphPad Prism v.6 (La Jolla, CA, USA).

### Acknowledgements
We are grateful to Andrea Buccarello and Fabrizia Ranieri for their technical assistance during experiments in rats, to Maddalena Tessari for technical assistance at the operating room. We thank Andrea Puglia for providing us the professional video camera tripod slider we used to stabilize our camera in operating room. We thank Prof. Emilio Macchi and Prof. Aderville Cabassi for the helpful discussion on the manuscript. We also thank Rhiannon Baird for careful English language revision of the manuscript. This work was supported by young research grant GR-2009-1530528 to M.M., the flagship program Nanomax miRNano to M.M. and the Horizon 2020 CUPIDO project GA 720834 to M.M.

### Author Contributions
G.R., S.R. and M.M. designed and performed the experiments. S.S., P.C.F. designed and performed modeling. L.F. designed and implemented the 2D-3D algorithm for heart contraction study. F.D.B. contributed to the modeling. G.P. and G.F. oversaw the project at clinical stage and performed TEE. M.M., M.G., F.P.L.M., L.F., G.R. analyzed and interpreted the data. All authors contributed in editing the final manuscript.

### Additional Information
**Supplementary information** accompanies this paper at http://www.nature.com/srep

**Competing Interests:** The authors declare no competing financial interests.

**How to cite this article**: Fassina, L. *et al.* Cardiac kinematic parameters computed from video of *in situ* beating heart. *Sci. Rep.* **7,** 46143; doi: 10.1038/srep46143 (2017).

**Publisher's note:** Springer Nature remains neutral with regard to jurisdictional claims in published maps and institutional affiliations.